# Computational Grids[*]

Ian Foster
Mathematics and Computer Science Division
Argonne National Laboratory
Argonne, IL 60439

Carl Kesselman
Information Sciences Institute
University of Southern California
Marina del Rey, CA 90292

In this introductory chapter, we lay the groundwork for the rest of the book by providing a more detailed picture of the expected purpose, shape, and architecture of future grid systems. We structure the chapter in terms of six questions that we believe are central to this discussion: Why do we need computational grids? What types of applications will grids be used for? Who will use grids? How will grids be used? What is involved in building a grid? And, what problems must be solved to make grids commonplace? We provide an overview of each of these issues here, referring to subsequent chapters for more detailed discussion.

## 1  Reasons for Computational Grids

Why do we need computational grids? Computational approaches to problem solving have proven their worth in almost every field of human endeavor. Computers are used for modeling and simulating complex scientific and engineering problems, diagnosing medical conditions, controlling industrial equipment, forecasting the weather, managing stock portfolios, and many other purposes. Yet, although there are certainly challenging problems that exceed our ability to solve them, computers are still used much less extensively than they could be. To pick just one example, university researchers make extensive use of computers when studying the impact of changes in land use on biodiversity, but city planners selecting routes for new roads or planning new zoning ordinances do not. Yet it is local decisions such as these that, ultimately, shape our future.

There are a variety of reasons for this relative lack of use of computational problem-solving methods, including lack of appropriate education and tools. But one important factor is that the average computing environment remains inadequate for such computationally sophisticated purposes. While today's PC is faster than the Cray supercomputer of 10 years ago, it is still far from adequate for predicting the outcome of complex actions or selecting from among many choices. That, after all, is why supercomputers have continued to evolve.

---

[*]Reprinted by permission of Morgan Kaufmann Publishers from *The Grid: Blueprint for a Future Computing Infrastructure*, I. Foster and C. Kesselman (Eds), 1998.

## 1.1 Increasing Delivered Computation

We believe that the opportunity exists to provide users—whether city planners, engineers, or scientists—with substantially more computational power: an increase of three orders of magnitude within five years, and five orders of magnitude within a decade. These dramatic increases will be achieved by innovations in a wide range of areas:

1. *Technology improvement:* Evolutionary changes in VLSI technology and microprocessor architecture can be expected to result in a factor of 10 increase in computational capabilities in the next five years, and a factor of 100 increase in the next ten.

2. *Increase in demand-driven access to computational power:* Many applications have only episodic requirements for substantial computational resources. For example, a medical diagnosis system may be run only when a cardiogram is performed, a stockmarket simulation only when a user recomputes retirement benefits, or a seismic simulation only after a major earthquake. If mechanisms are in place to allow reliable, instantaneous, and transparent access to high-end resources, then from the perspective of these applications it is as if those resources are dedicated to them. Given the existence of multiteraFLOPS systems, an increase in apparent computational power of three or more orders of magnitude is feasible.

3. *Increased utilization of idle capacity:* Most low-end computers (PCs and workstations) are often idle: various studies report utilizations of around 30% in academic and commercial environments [47], [21]. Utilization can be increased by a factor of two, even for parallel programs [4], without impinging significantly on productivity. The benefit to individual users can be substantially greater: factors of 100 or 1,000 increase in peak computational capacity have been reported [41], [75].

4. *Greater sharing of computational results:* The daily weather forecast involves perhaps $10^{14}$ numerical operations. If we assume that the forecast is of benefit to $10^7$ people, we have $10^{21}$ effective operations—comparable to the computation performed each day on all the world's PCs. Few other computational results or facilities are shared so effectively today, but they may be in the future as other scientific communities adopt a "big science" approach to computation. The key to more sharing may be the development of collaboratories: "... center[s] without walls, in which the nation's researchers can perform their research without regard to geographical location—interacting with colleagues, accessing instrumentation, sharing data and computational resources, and accessing information in digital libraries" [48].

5. *New problem-solving techniques and tools:* A variety of approaches can improve the efficiency or ease with which computation is applied to problem solving. For example, network-enabled solvers [17], [11] allow users to invoke advanced numerical solution methods without having to install sophisticated software. Teleimmersion techniques [50] facilitate the sharing of computational results by supporting collaborative steering of simulations and exploration of data sets.

Underlying each of these advances is the synergistic use of high-performance networking, computing, and advanced software to provide access to advanced computational capabilities, regardless of the location of users and resources.

## 1.2 Definition of Computational Grids

The current status of computation is analogous in some respects to that of electricity around 1910. At that time, electric power generation was possible, and new devices were being devised that depended on electric power, but the need for each user to build and operate a new generator hindered use. The truly revolutionary development was not, in fact, electricity, but the electric power grid and the associated transmission and distribution technologies. Together, these developments provided reliable, low-cost access to a standardized service, with the result that power—which for most of human history has been accessible only in crude and not especially portable forms (human effort, horses, water power, steam engines, candles)—became universally accessible. By allowing both individuals and industries to take for granted the availability of cheap, reliable power, the electric power grid made possible both new devices and the new industries that manufactured them.

By analogy, we adopt the term *computational grid* for the infrastructure that will enable the increases in computation discussed above. A computational grid is a hardware and software infrastructure that provides dependable, consistent, pervasive, and inexpensive access to high-end computational capabilities.

We talk about an *infrastructure* because a computational grid is concerned, above all, with large-scale pooling of resources, whether compute cycles, data, sensors, or people. Such pooling requires significant hardware infrastructure to achieve the necessary interconnections and software infrastructure to monitor and control the resulting ensemble. In the rest of this chapter, and throughout the book, we discuss in detail the nature of this infrastructure.

The need for *dependable* service is fundamental. Users require assurances that they will receive predictable, sustained, and often high levels of performance from the diverse components that constitute the grid; in the absence of such assurances, applications will not be written or used. The performance characteristics that are of interest will vary widely from application to application, but may include network bandwidth, latency, jitter, computer power, software services, security, and reliability.

The need for *consistency* of service is a second fundamental concern. As with electric power, we need standard services, accessible via standard interfaces, and operating within standard parameters. Without such standards, application development and pervasive use are impractical. A significant challenge when developing standards is to encapsulate heterogeneity without compromising high-performance execution.

*Pervasive* access allows us to count on services always being available, within whatever environment we expect to move. Pervasiveness does not imply that resources are everywhere or are universally accessible. We cannot access electric power in a new home until wire has been laid and an account established with the local utility; computational grids will have similarly circumscribed availability and controlled access. However, we will be able to count on universal access within the confines of whatever environment the grid is designed to support.

Finally, an infrastructure must offer *inexpensive* (relative to income) access if it is to be broadly accepted and used. Homeowners and industrialists both make use of remote billion-dollar power plants on a daily basis because the cost to them is reasonable. A computational grid must achieve similarly attractive economics.

It is the combination of dependability, consistency, and pervasiveness that will cause computational grids to have a transforming effect on how computation is performed and used. By increasing the set of capabilities that can be taken for granted to the extent that they are noticed only by their absence, grids allow new tools to be developed and widely deployed. Much as pervasive access to bitmapped displays changed our baseline assumptions for the design of application interfaces, computational grids can fundamentally change the way we think about computation and resources.

## 1.3 The Impact of Grids

The history of network computing shows that orders-of-magnitude improvements in underlying technology invariably enable revolutionary, often unanticipated, applications of that technology, which in turn motivate further technological improvements. As a result, our view of network computing has undergone repeated transformations over the past 40 years.

There is considerable evidence that another such revolution is imminent. The capabilities of both computers and networks continue to increase dramatically. Ten years of research on metacomputing has created a solid base of experience in new applications that couple high-speed networking and computing. The time seems ripe for a transition from the heroic days of metacomputing to more integrated computational grids with dependable and pervasive computational capabilities and consistent interfaces. In such grids, today's metacomputing applications will be routine, and programmers will be able to explore a new generation of yet more interesting applications that leverage teraFLOP computers and petabyte storage systems interconnected by gigabit networks. We present two simple examples to illustrate how grid functionality may transform different aspects of our lives.

Today's home finance software packages leverage the pervasive availability of communication technologies such as modems, Internet service providers, and the Web to integrate up-to-date stock prices obtained from remote services into local portfolio value calculations. However, the actual computations performed on this data are relatively simple. In tomorrow's grid environment, we can imagine individuals making stock-purchasing decisions on the basis of detailed Monte Carlo analyses of future asset value, performed on remote teraFLOP computers. The instantaneous use of three orders of magnitude more computing power than today will go unnoticed by prospective retirees, but their lives will be different because of more accurate information.

Today, citizen groups evaluating a proposed new urban development must study uninspiring blueprints or perspective drawings at city hall. A computational grid will allow them to call on powerful graphics computers and databases to transform the architect's plans into realistic virtual reality depictions and to explore such design issues as energy consumption, lighting efficiency, or sound quality. Meeting online to walk through and discuss the impact of the new development on their community, they can arrive at better urban design and hence improved quality of life. Virtual reality-based simulation models of Los Angeles, produced by William Jepson, and the walkthrough model of Soda Hall at the University of California–Berkeley, constructed by Carlo Seguin and his colleagues, are interesting exemplars of this use of computing [9].

## 1.4 Electric Power Grids

We conclude this section by reviewing briefly some salient features of the computational grid's namesake. The electric power grid is remarkable in terms of its construction and function, which together make it one of the technological marvels of the 20th century. Within large geographical regions (e.g., North America), it forms essentially a single entity that provides power to billions of devices, in a relatively efficient, low-cost, and reliable fashion. The North American grid alone links more than ten thousand generators with billions of outlets via a complex web of physical connections and trading mechanisms [12]. The components from which the grid is constructed are highly heterogeneous in terms of their physical characteristics and are owned and operated by different organizations. Consumers differ significantly in terms of the amount of power they consume, the service guarantees they require, and the amount they are prepared to pay.

Analogies are dangerous things, and electricity is certainly very different from computation in many respects. Nevertheless, the following aspects of the power grid seem particularly relevant to the current discussion.

**Importance of Economics**

The role and structure of the power grid are driven to a large extent by economic factors. Oil- and coal-fired generators have significant economies of scale. A power company must be able to call upon reserve capacity equal to its largest generator in case that generator fails; interconnections between regions allow for sharing of such reserve capacity, as well as enabling trading of excess power. The impact of economic factors on computational grids is not well understood [34]. Where and when are there economies of scale to be obtained in computational capabilities? Might economic factors lead us away from today's model of a "computer on every desktop"? We note an intriguing development. Recent advances in power generation technology (e.g., small gas turbines) and the deregulation of the power industry are leading some analysts to look to the Internet for lessons regarding the future evolution of the electric power grid!

**Importance of Politics**

The developers of large-scale grids tell us that their success depended on regulatory, political, and institutional developments as much as on technical innovation [12]. This lesson should be taken to heart by developers of future computational grids.

**Complexity of Control**

The principal technical challenges in power grids—once technology issues relating to efficient generation and high-voltage transmission had been overcome—relate to the management of a complex ensemble in which changes at a single location can have far-reaching consequences [12]. Hence, we find that the power grid includes a sophisticated infrastructure for monitoring, management, and control. Again, there appear to be many parallels between this control problem and the problem of providing performance guarantees in large-scale, dynamic, and heterogeneous computational grid environments.

## 2 Grid Applications

What types of applications will grids be used for? Building on experiences in gigabit testbeds [42], [59], the I-WAY network [19], and other experimental systems, we have identified five major application classes for computational grids, listed in Table 1 and described briefly in this section. More details about applications and their technical requirements are provided in the referenced chapters.

### 2.1 Distributed Supercomputing

Distributed supercomputing applications use grids to aggregate substantial computational resources in order to tackle problems that cannot be solved on a single system. Depending on the grid on which we are working (see Section 3), these aggregated resources might comprise the majority of the supercomputers in the country or simply all of the workstations within a company. Here are some contemporary examples:

- Distributed interactive simulation (DIS) is a technique used for training and planning in the military. Realistic scenarios may involve hundreds of thousands of entities, each with potentially complex behavior patterns. Yet even the largest current supercomputers can handle at most 20,000 entities. In recent work, researchers at the California Institute of Technology have shown how multiple supercomputers can be coupled to achieve record-breaking levels of performance.

- The accurate simulation of complex physical processes can require high spatial and temporal resolution in order to resolve fine-scale detail. Coupled supercomputers can be used in such situations to overcome resolution barriers and hence to obtain qualitatively new scientific results. Although high latencies can pose significant obstacles, coupled supercomputers have been used successfully in cosmology [54], high-resolution ab initio computational chemistry computations [52], and climate modeling [45].

| Category | Examples | Characteristics |
|---|---|---|
| Distributed supercomputing | DIS<br>Stellar dynamics<br>Ab initio chemistry | Very large problems needing lots of CPU, memory, etc. |
| High throughput | Chip design<br>Parameter studies<br>Cryptographic problems | Harness many otherwise idle resources to increase aggregate throughput |
| On demand | Medical instrumentation<br>Network-enabled solvers<br>Cloud detection | Remote resources integrated with local computation, often for bounded amount of time |
| Data intensive | Sky survey<br>Physics data<br>Data assimilation | Synthesis of new information from many or large data sources |
| Collaborative | Collaborative design<br>Data exploration<br>Education | Support communication or collaborative work between multiple participants |

Table 1: Five major classes of grid applications.

Challenging issues from a grid architecture perspective include the need to coschedule what are often scarce and expensive resources, the scalability of protocols and algorithms to tens or hundreds of thousands of nodes, latency-tolerant algorithms, and achieving and maintaining high levels of performance across heterogeneous systems.

## 2.2 High-Throughput Computing

In high-throughput computing, the grid is used to schedule large numbers of loosely coupled or independent tasks, with the goal of putting unused processor cycles (often from idle workstations) to work. The result may be, as in distributed supercomputing, the focusing of available resources on a single problem, but the quasi-independent nature of the tasks involved leads to very different types of problems and problem-solving methods. Here are some examples:

- Platform Computing Corporation reports that the microprocessor manufacturer Advanced Micro Devices used high-throughput computing techniques to exploit over a thousand computers during the peak design phases of their K6 and K7 microprocessors. These computers are located on the desktops of AMD engineers at a number of AMD sites and were used for design verification only when not in use by engineers.

- The Condor system from the University of Wisconsin is used to manage pools of hundreds of workstations at universities and laboratories around the world [41]. These resources have been used for studies as diverse as molecular simulations of liquid crystals, studies of ground-penetrating radar, and the design of diesel engines.

- More loosely organized efforts have harnessed tens of thousands of computers distributed worldwide to tackle hard cryptographic problems [40].

## 2.3 On-Demand Computing

On-demand applications use grid capabilities to meet short-term requirements for resources that cannot be cost-effectively or conveniently located locally. These resources may be computation, software, data repositories, specialized sensors, and so on. In contrast to distributed supercomputing applications, these applications are often driven by cost-performance concerns rather than absolute performance. For example:

- The NEOS [17] and NetSolve [11] network-enhanced numerical solver systems allow users to couple remote software and resources into desktop applications, dispatching to remote servers calculations that are computationally demanding or that require specialized software.

- A computer-enhanced MRI machine and scanning tunneling microscope (STM) developed at the National Center for Supercomputing Applications use supercomputers to achieve realtime image processing [57], [58]. The result is a significant enhancement in the ability to understand what we are seeing and, in the case of the microscope, to steer the instrument.

- A system developed at the Aerospace Corporation for processing of data from meteorological satellites uses dynamically acquired supercomputer resources to deliver the results of a cloud detection algorithm to remote meteorologists in quasi real time [38].

The challenging issues in on-demand applications derive primarily from the dynamic nature of resource requirements and the potentially large populations of users and resources. These issues include resource location, scheduling, code management, configuration, fault tolerance, security, and payment mechanisms.

## 2.4 Data-Intensive Computing

In data-intensive applications, the focus is on synthesizing new information from data that is maintained in geographically distributed repositories, digital libraries, and databases. This synthesis process is often computationally and communication intensive as well.

- Future high-energy physics experiments will generate terabytes of data per day, or around a petabyte per year. The complex queries used to detect "interesting" events may need to access large fractions of this data [43]. The scientific collaborators who will access this data are widely distributed, and hence the data systems in which data is placed are likely to be distributed as well.

- The Digital Sky Survey will, ultimately, make many terabytes of astronomical photographic data available in numerous network-accessible databases. This facility enables new approaches to astronomical research based on distributed analysis, assuming that appropriate computational grid facilities exist.

- Modern meteorological forecasting systems make extensive use of data assimilation to incorporate remote satellite observations. The complete process involves the movement and processing of many gigabytes of data.

Challenging issues in data-intensive applications are the scheduling and configuration of complex, high-volume data flows through multiple levels of hierarchy.

## 2.5 Collaborative Computing

Collaborative applications are concerned primarily with enabling and enhancing human-to-human interactions. Such applications are often structured in terms of a virtual shared space. Many collaborative applications are concerned with enabling the shared use of computational resources such as data archives and simulations; in this case, they also have characteristics of the other application classes just described. For example:

- The BoilerMaker system developed at Argonne National Laboratory allows multiple users to collaborate on the design of emission control systems in industrial incinerators [20]. The different users interact with each other and with a simulation of the incinerator.

- The CAVE5D system supports remote, collaborative exploration of large geophysical data sets and the models that generate them—for example, a coupled physical/biological model of the Chesapeake Bay [74].

- The NICE system developed at the University of Illinois at Chicago allows children to participate in the creation and maintenance of realistic virtual worlds, for entertainment and education [60].

Challenging aspects of collaborative applications from a grid architecture perspective are the real-time requirements imposed by human perceptual capabilities and the rich variety of interactions that can take place.

We conclude this section with three general observations. First, we note that even in this brief survey we see a tremendous variety of already successful applications. This rich set has been developed despite the significant difficulties faced by programmers developing grid applications in the absence of a mature grid infrastructure. As grids evolve, we expect the range and sophistication of applications to increase dramatically. Second, we observe that almost all of the applications demonstrate a tremendous appetite for computational resources (CPU, memory, disk, etc.) that cannot be met in a timely fashion by expected growth in single-system performance. This emphasizes the importance of grid technologies as a means of sharing computation as well as a data access and communication medium. Third, we see that many of the applications are interactive, or depend on tight synchronization with computational components, and hence depend on the availability of a grid infrastructure able to provide robust performance guarantees.

# 3 Grid Communities

Who will use grids? One approach to understanding computational grids is to consider the communities that they serve. Because grids are above all a mechanism for sharing resources, we ask, What groups of people will have sufficient incentive to invest in the infrastructure required to enable sharing, and what resources will these communities want to share?

One perspective on these questions holds that the benefits of sharing will almost always outweigh the costs and, hence, that we will see grids that link large communities with few common interests, within which resource sharing will extend to individual PCs and workstations. If we compare a computational grid to an electric power grid, then in this view, the grid is quasi-universal, and every user has the potential to act as a cogenerator. Skeptics respond that the technical and political costs of sharing resources will rarely outweigh the benefits, especially when coupling must cross institutional boundaries. Hence, they argue that resources will be shared only when there is considerable incentive to do so: because the resource is expensive, or scarce, or because sharing enables human interactions that are otherwise difficult to achieve. In this view, grids will be specialized, designed to support specific user communities with specific goals.

Rather than take a particular position on how grids will evolve, we propose what we see as four plausible scenarios, each serving a different community. Future grids will probably include elements of all four.

## 3.1 Government

The first community that we consider comprises the relatively small number—thousands or perhaps tens of thousands—of officials, planners, and scientists concerned with problems traditionally assigned to national government, such as disaster response, national defense, and long-term research and planning. There can be significant advantage to applying the collective power of the nation's fastest computers, data archives, and intellect to the solution of these problems. Hence, we envision a grid that uses the fastest networks to couple relatively small numbers of high-end resources across the nation—perhaps tens of teraFLOP computers, petabytes of storage, hundreds of sites, thousands of smaller systems—for two principal purposes:

1. To provide a "strategic computing reserve," allowing substantial computing resources to be applied to large problems in times of crisis, such as to plan responses to a major environmental disaster, earthquake, or terrorist attack

2. To act as a "national collaboratory," supporting collaborative investigations of complex scientific and engineering problems, such as global change, space station design, and environmental cleanup

An important secondary benefit of this high-end national supercomputing grid is to support resource trading between the various operators of high-end resources, hence increasing the efficiency with which those resources are used.

This *national grid* is distinguished by its need to integrate diverse high-end (and hence complex) resources, the strategic importance of its overall mission, and the diversity of competing interests that must be balanced when allocating resources.

## 3.2 A Health Maintenance Organization

In our second example, the community supported by the grid comprises administrators and medical personnel located at a small number of hospitals within a metropolitan area. The resources to be shared are a small number of high-end computers, hundreds of workstations, administrative databases, medical image archives, and specialized instruments such as MRI machines, CAT scanners, and cardioangiography devices. The coupling of these resources into an integrated grid enables a wide range of new, computationally enhanced applications: desktop tools that use centralized supercomputer resources to run computer-aided diagnosis procedures on mammograms or to search centralized medical image archives for similar cases; life-critical applications such as telerobotic surgery and remote cardiac monitoring and analysis; auditing software that uses the many workstations across the hospital to run fraud detection algorithms on financial records; and research software that uses supercomputers and idle workstations for epidemiological research. Each of these applications exists today in research laboratories, but has rarely been deployed in ordinary hospitals because of the high cost of computation.

This *private grid* is distinguished by its relatively small scale, central management, and common purpose on the one hand, and on the other hand by the complexity inherent in using common infrastructure for both life-critical applications and less reliability-sensitive purposes and by the need to integrate low-cost commodity technologies. We can expect grids with similar characteristics to be useful in many institutions.

### 3.3 A Materials Science Collaboratory

The community in our third example is a group of scientists who operate and use a variety of instruments, such as electron microscopes, particle accelerators, and X-ray sources, for the characterization of materials. This community is fluid and highly distributed, comprising many hundreds of university researchers and students from around the world, in addition to the operators of the various instruments (tens of instruments, at perhaps ten centers). The resources that are being shared include the instruments themselves, data archives containing the collective knowledge of this community, sophisticated analysis software developed by different groups, and various supercomputers used for analysis. Applications enabled by this grid include remote operation of instruments, collaborative analysis, and supercomputer-based online analysis.

This *virtual grid* is characterized by a strong unifying focus and relatively narrow goals on the one hand, and on the other hand by dynamic membership, a lack of central control, and a frequent need to coexist with other uses of the same resources. We can imagine similar grids arising to meet the needs of a variety of multi-institutional research groups and multicompany virtual teams created to pursue long- or short-term goals.

### 3.4 Computational Market Economy

The fourth community that we consider comprises the participants in a broad-based market economy for computational services. This is a potentially enormous community with no connections beyond the usual market-oriented relationships. We can expect participants to include consumers, with their diverse needs and interests; providers of specialized services, such as financial modeling, graphics rendering, and interactive gaming; providers of compute resources; network providers, who contract to provide certain levels of network service; and various other entities such as banks and licensing organizations.

This *public grid* is in some respects the most intriguing of the four scenarios considered here, but is also the least concrete. One area of uncertainty concerns the extent to which the average consumer will also act as a producer of computational resources. The answer to this question seems to depend on two issues. Will applications emerge that can exploit loosely coupled computational resources? And, will owners of resources be motivated to contribute resources? To date, large-scale activity in this area has been limited to fairly esoteric computations—such as searching for prime numbers, breaking cryptographic codes [40], or detecting extraterrestrial communications [64]—with the benefit to the individuals being the fun of participating and the potential momentary fame if their computer solves the problem in question.

We conclude this section by noting that, in our view, each of these scenarios seems quite feasible; indeed, substantial prototypes have been created for each of the grids that we describe. Hence, we expect to see not just one single computational grid, but rather many grids, each serving a different community with its own requirements and objectives. Just which grids will evolve depends critically on three issues: the evolving economics of computing and networking, and the services that these physical infrastructure elements are used to provide; the institutional, regulatory, and political frameworks within which grids may develop; and, above all, the emergence of applications able to motivate users to invest in and use grid technologies.

## 4 Using Grids

How will grids be used? In metacomputing experiments conducted to date, users have been "heroic" programmers, willing to spend large amounts of time programming complex systems at a low level.

| Class | Purpose | Makes use of | Concerns |
|---|---|---|---|
| End users | Solve problems | Applications | Transparency, performance |
| Application developers | Develop applications | Programming models, tools | Ease of use, performance |
| Tool developers | Develop tools, programming models | Grid services | Adaptivity, exposure of performance, security |
| Grid developers | Provide basic grid services | Local system services | Local simplicity, connectivity, security |
| System administrators | Manage grid resources | Management tools | Balancing local and global concerns |

Table 2: Classes of grid users.

The resulting applications have provided compelling demonstrations of what might be, but in most cases are too expensive, unreliable, insecure, and fragile to be considered suitable for general use.

For grids to become truly useful, we need to take a significant step forward in grid programming, moving from the equivalent of assembly language to high-level languages, from one-off libraries to application toolkits, and from hand-crafted codes to shrink-wrapped applications. These goals are familiar to us from conventional programming, but in a grid environment we are faced with the additional difficulties associated with wide area operation—in particular, the need for grid applications to adapt to changes in resource properties in order to meet performance requirements. As in conventional computing, an important step toward the realization of these goals is the development of standards for applications, programming models, tools, and services, so that a division of labor can be achieved between the users and developers of different types of components.

We structure our discussion of grid tools and programming in terms of the classification illustrated in Table 2. At the lowest level, we have *grid developers*—the designers and implementors of what we might call the "Grid Protocol," by analogy with the Internet Protocol that provides the lowest-level services in the Internet—who provide the basic services required to construct a grid. Above this, we have *tool developers*, who use grid services to construct programming models and associated tools, layering higher-level services and abstractions on top of the more fundamental services provided by the grid architecture. *Application developers*, in turn, build on these programming models, tools, and services to construct grid-enabled applications for *end users* who, ideally, can use these applications without being concerned with the fact that they are operating in a grid environment. A fifth class of users, *system administrators*, is responsible for managing grid components. We now examine this model in more detail.

## 4.1 Grid Developers

A very small group of grid developers are responsible for implementing the basic services referred to above. We discuss the concerns encountered at this level in Section 5.

## 4.2 Tool Developers

Our second group of users are the developers of the tools, compilers, libraries, and so on that implement the programming models and services used by application developers. Today's small population of grid tool developers (e.g., the developers of Condor [41], Nimrod [1], NEOS [17], NetSolve [11], Horus [68], grid-enabled implementations of the Message Passing Interface (MPI) [27], and CAVERN [39]) must build their tools on a very narrow foundation, comprising little more than the Internet Protocol. We envision that future grid systems will provide a richer set of basic services, hence making it possible to build more sophisticated and robust tools. We discuss the nature and implementation of those basic services in Section 5; briefly, they comprise versions of those services that have proven effective on today's end systems and clusters, such as authentication, process management, data access, and communication, plus new services that address specific concerns of the grid environment, such as resource location, information, fault detection, security, and electronic payment.

Tool developers must use these basic services to provide efficient implementations of the programming models that will be used by application developers. In constructing these translations, the tool developer must be concerned not only with translating the existing model to the grid environment, but also with revealing to the programmer those aspects of the grid environment that impact performance. For example, a grid-enabled MPI [27] can seek to adapt the MPI model for grid execution by incorporating specialized techniques for point-to-point and collective communication in highly heterogeneous environments; implementations of collective operations might use multicast protocols and adapt a combining tree structure in response to changing network loads. It should probably also extend the MPI model to provide programmers with access to resource location services, information about grid topology, and group communication protocols.

### 4.3 Application Developers

Our third class of users comprises those who construct grid-enabled applications and components. Today, these programmers write applications in what is, in effect, an assembly language: explicit calls to the Internet Protocol's User Datagram Protocol (UDP) or Transmission Control Protocol (TCP), explicit or no management of failure, hard-coded configuration decisions for specific computing systems, and so on. We are far removed from the portable, efficient, high-level languages that are used to develop sequential programs, and the advanced services that programmers can rely upon when using these languages, such as dynamic memory management and high-level I/O libraries.

Future grids will need to address the needs of application developers in two ways. They must provide *programming models* (supported by languages, libraries, and tools) that are appropriate for grid environments and a range of *services* (for security, fault detection, resource management, data access, communication, etc.) that programmers can call upon when developing applications.

The purpose of both programming models and services is to simplify thinking about and implementing complex algorithmic structures, by providing a set of abstractions that hide details unrelated to the application, while exposing design decisions that have a significant impact on program performance or correctness. In sequential programming, commonly used programming models provide us with abstractions such as subroutines and scoping; in parallel programming, we have threads and condition variables (in shared-memory parallelism), message passing, distributed arrays, and single-assignment variables. Associated services ensure that resources are allocated to processes in a reasonable fashion, provide convenient abstractions for tertiary storage, and so forth.

There is no consensus on what programming model is appropriate for a grid environment, although it seems clear that many models will be used. Table 3 summarizes some of the models that have been proposed; new models will emerge as our understanding of grid programming evolves.

As Table 3 makes clear, one approach to grid programming is to adapt models that have already proved successful in sequential or parallel environments. For example, a grid-enabled distributed shared-memory (DSM) system would support a shared-memory programming model in a grid environment, allowing programmers to specify parallelism in terms of threads and shared-memory operations. Similarly, a grid-enabled MPI would extend the popular message-passing model [27], and a grid-enabled file system would permit remote files to be accessed via the standard UNIX application programming interface (API) [66]. These approaches have the advantage of potentially allowing existing applications to be reused unchanged, but can introduce significant performance problems if the models in question do not adapt well to high-latency, dynamic, heterogeneous grid environments.

| Model | Examples | Pros | Cons |
| --- | --- | --- | --- |
| Datagram/stream communication | UDP, TCP, Multicast | Low overhead | Low level |
| Shared memory, multithreading | POSIX Threads DSM | High level | Scalability |
| Data parallelism | HPF, HPC++ | Automatic parallelization | Restricted applicability |
| Message passing | MPI, PVM | High performance | Low level |
| Object-oriented | CORBA, DCOM, Java RMI | Support for large-system design | Performance |
| Remote procedure call | DCE, ONC | Simplicity | Restricted applicability |
| High throughput | Condor, LSF, Nimrod | Ease of use | Restricted applicability |
| Group ordered | Isis, Totem | Robustness | Performance, scalability |
| Agents | Aglets, Telescript | Flexibility | Performance, robustness |

Table 3: Potential grid programming models and their advantages and disadvantages.

Another approach is to build on technologies that have proven effective in distributed computing, such as Remote Procedure Call (RPC) or related object-based techniques such as the Common Object Request Broker Architecture (CORBA). These technologies have significant software engineering advantages, because their encapsulation properties facilitate the modular construction of programs and the reuse of existing components. However, it remains to be seen whether these models can support performance-focused, complex applications such as teleimmersion or the construction of dynamic computations that span hundreds or thousands of processors.

The grid environment can also motivate new programming models and services. For example, high-throughput computing systems, as exemplified by Condor [41] and Nimrod [1], support problem-solving methods such as parameter studies in which complex problems are partitioned into many independent tasks. Group-ordered communication systems represent another model that is important in dynamic, unpredictable grid environments; they provide services for managing groups of processes and for delivering messages reliably to group members. Agent-based programming models represent another approach apparently well suited to grid environments; here, programs are constructed as independent entities that roam the network searching for data or performing other tasks on behalf of a user.

A wide range of new services can be expected to arise in grid environments to support the development of more complex grid applications. In addition to grid analogs of conventional services such as file systems, we will see new services for resource discovery, resource brokering, electronic payments, licensing, fault tolerance, specification of use conditions, configuration, adaptation, and distributed system management, to name just a few.

### 4.4 End Users

Most grid users, like most users of computers or networks today, will not write programs. Instead, they will use grid-enabled applications that make use of grid resources and services. These applications may be chemistry packages or environmental models that use grid resources for computing or data; problem-solving packages that help set up parameter study experiments [1]; mathematical packages augmented with calls to network-enabled solvers [17], [11]; or collaborative engineering packages that allow geographically separated users to cooperate on the design of complex systems.

End users typically place stringent requirements on their tools, in terms of reliability, predictability, confidentiality, and usability. The construction of applications that can meet these requirements in complex grid environments represents a major research and engineering challenge.

### 4.5 System Administrators

The final group of users that we consider are the system administrators who must manage the infrastructure on which computational grids operate. This task is complicated by the high degree of sharing that grids are designed to make possible. The user communities and resources associated with a particular grid will frequently span multiple administrative domains, and new services will arise—such as accounting and resource brokering—that require distributed management. Furthermore, individual resources may participate in several different grids, each with its own particular user community, access policies, and so on. For a grid to be effective, each participating resource must be administered so as to strike an appropriate balance between local policy requirements and the needs of the larger grid community. This problem has a significant political dimension, but new technical solutions are also required.

The Internet experience suggests that two keys to scalability when administering large distributed systems are to decentralize administration and to automate trans-site issues. For example, names and routes are administered locally, while essential trans-site services such as route discovery and name resolution are automated. Grids will require a new generation of tools for automatically monitoring and managing many tasks that are currently handled manually.

New administration issues that arise in grids include establishing, monitoring, and enforcing local policies in situations where the set of users may be large and dynamic; negotiating policy with other sites and users; accounting and payment mechanisms; and the establishment and management of markets and other resource-trading mechanisms. There are interesting parallels between these problems and management issues that arise in the electric power and banking industries 114, [31], [28].

## 5 Grid Architecture

What is involved in building a grid? To address this question, we adopt a system architect's perspective and examine the organization of the software infrastructure required to support the grid users, applications, and services discussed in the preceding sections.

As noted above, computational grids will be created to serve different communities with widely varying characteristics and requirements. Hence, it seems unlikely that we will see a single grid architecture. However, we do believe that we can identify basic services that most grids will provide, with different grids adopting different approaches to the realization of these services.

One major driver for the techniques used to implement grid services is scale. Computational infrastructure, like other infrastructures, is fractal, or self-similar at different scales. We have networks between countries, organizations, clusters, and computers; between components of a computer; and even within a single component. However, at different scales, we often operate in different physical, economic, and political regimes. For example, the access control solutions used for a laptop computer's system bus are probably not appropriate for a trans-Pacific cable.

| Comp. model | I/O model | Resource manag. | Security |
|---|---|---|---|
| **Endsystem:** | | | |
| Multithreading, automatic parallelization, | Local I/O, disk-striping | Process creation OS signal delivery OS scheduling | OS kernel, hardware |
| **Cluster** (increased scale, reduced integration): | | | |
| Synchronous communication, distributed shared memory | Parallel I/O (e.g., MPI-IO), file systems | Parallel process creation, gang scheduling, OS-level signal propagation | Shared security databases |
| **Intranet** (heterogeneity, separate administration, lack of global knowledge): | | | |
| Client/server, loosely synchronous: pipelines, coupling manager/worker | Distributed file systems (DFS, HPSS), databases | Resource discovery, signal distribution networks, high throughput | Network security (Kerberos) |
| **Internet** (lack of centralized control, geographical distribution, intl. issues): | | | |
| Collaborative systems, remote control, data mining | Remote file access, digital libraries, data warehouses | Brokers, trading, mobile code negotiation | Trust delegation, public key, sandboxes |

Table 4: Computer systems operating at different scales.

In this section, we adopt scale as the major dimension for comparison. We consider four types of systems, of increasing scale and complexity, asking two questions for each: What new concerns does this increase in scale introduce? And how do these new concerns influence how we provide basic services? These system types are as follows (see also Table 4):

1. The *end system* provides the best model we have for what it means to compute, because it is here that most research and development efforts have focused in the past four decades.

2. The *cluster* introduces new issues of parallelism and distributed management, albeit of homogeneous systems.

3. The *intranet* introduces the additional issues of heterogeneity and geographical distribution.

4. The *internet* introduces issues associated with a lack of centralized control.

An important secondary driver for architectural solutions is the performance requirements of the grid. Stringent performance requirements amplify the effect of scale because they make it harder to hide heterogeneity. For example, if performance is not a big concern, it is straightforward to extend UNIX file I/O to support access to remote files, perhaps via a HyperText Transport Protocol (HTTP) gateway [66]. However, if performance is critical, remote access may require quite different mechanisms—such as parallel transfers over a striped network from a remote parallel file system to a local parallel computer—that are not easily expressed in terms of UNIX file I/O semantics. Hence, a high-performance wide area grid may need to adopt quite different solutions to data access problems. In the following, we assume that we are dealing with high-performance systems; systems with lower performance requirements are generally simpler.

## 5.1 Basic Services

We start our discussion of architecture by reviewing the basic services provided on conventional computers. We do so because we believe that, in the absence of strong evidence to the contrary, services that have been developed and proven effective in several decades of conventional computing will also be desirable in computational grids. Grid environments also require additional services, but we claim that, to a significant extent, grid development will be concerned with extending familiar capabilities to the more complex wide area environment.

Our purpose in this subsection is not to provide a detailed exposition of well-known ideas but rather to establish a vocabulary for subsequent discussion. We assume that we are discussing a generic modern computing system, and hence refrain from prefixing each statement with "in general," "typically," and the like. Individual systems will, of course, differ from the generic systems described here, sometimes in interesting and important ways.

The first step in a computation that involves shared resources is an *authentication* process, designed to establish the identity of the user. A subsequent *authorization* process establishes the right of the user to create entities called *processes*. A process comprises one or more *threads* of control, created for either concurrency or parallelism, and executing within a *shared address space*. A process can also *communicate* with other processes via a variety of abstractions, including shared memory (with semaphores or locks), pipes, and protocols such as TCP/IP.

A user (or process acting on behalf of a user) can *control* the activities in another process— for example, to suspend, resume, or terminate its execution. This control is achieved by means of asynchronously delivered *signals*.

A process acts on behalf of its creator to *acquire resources*, by executing instructions, occupying memory, reading and writing disks, sending and receiving messages, and so on. The ability of a process to acquire resources is limited by underlying authorization mechanisms, which implement a system's *resource allocation policy*, taking into account the user's identity, prior resource consumption, and/or other criteria. *Scheduling* mechanisms in the underlying system deal with competing demands for resources and may also (for example, in realtime systems) support user requests for performance guarantees.

Underlying *accounting* mechanisms keep track of resource allocations and consumption, and *payment* mechanisms may be provided to translate resource consumption into some common currency. The underlying system will also provide *protection* mechanisms to ensure that one user's computation does not interfere with another's.

Other services provide abstractions for secondary storage. Of these, *virtual memory* is implicit, extending the shared address space abstraction already noted; *file systems* and *databases* are more explicit representations of secondary storage.

## 5.2 End Systems

Individual end systems—computers, storage systems, sensors, and other devices—are characterized by relatively small scale and a high degree of homogeneity and integration. There are typically just a few tens of components (processors, disks, etc.), these components are mostly of the same type, and the components and the software that controls them have been co-designed to simplify management and use and to maximize performance. (Specialized devices such as scientific instruments may be more significantly complex, with potentially thousands of internal components, of which hundreds may be visible externally.)

Such end systems represent the simplest, and most intensively studied, environment in which to provide the services listed above. The principal challenges facing developers of future systems of this type relate to changing computer architectures (in particular, parallel architectures) and the need to integrate end systems more fully into clusters, intranets, and internets.

**State of the Art**

The software architectures used in conventional end systems are well known [61]. Basic services are provided by a privileged operating system, which has absolute control over the resources of the computer. This operating system handles authentication and mediates user process requests to acquire resources, communicate with other processes, access files, and so on. The integrated nature of the hardware and operating system allows high-performance implementations of important functions such as virtual memory and I/O.

Programmers develop applications for these end systems by using a variety of high-level languages and tools. A high degree of integration between processor architecture, memory system, and compiler means that high performance can often be achieved with relatively little programmer effort.

**Future Directions**

A significant deficiency of most end-system architectures is that they lack features necessary for integration into larger clusters, intranets, and internets. Much current research and development is concerned with evolving system end architectures in directions relevant to future computational grids. To list just three: Operating systems are evolving to support operation in clustered environments, in which services are distributed over multiple networked computers, rather than replicated on every processor [3], [65]. A second important trend is toward a greater integration of end systems (computers, disks, etc.) with networks, with the goal of reducing the overheads incurred at network interfaces and hence increasing communication rates [22], [35]. Finally, support for mobile code is starting to appear, in the form of authentication schemes, secure execution environments for downloaded code ("sandboxes"), and so on [32], [72], [71], [44].

The net effect of these various developments seems likely to be to reduce the currently sharp boundaries between end system, cluster, and intranet/internet, with the result that individual end systems will more fully embrace remote computation, as producers and/or consumers.

## 5.3 Clusters

The second class of systems that we consider is the cluster, or network of workstations: a collection of computers connected by a high-speed local area network and designed to be used as an integrated computing or data processing resource. A cluster, like an individual end system, is a homogeneous entity—its constituent systems differ primarily in configuration, not basic architecture—and is controlled by a single administrative entity who has complete control over each end system. The two principal complicating factors that the cluster introduces are as follows:

1. *Increased physical scale*: A cluster may comprise several hundred or thousand processors, with the result that alternative algorithms are needed for certain resource management and control functions.

2. *Reduced integration*: A desire to construct clusters from commodity parts means that clusters are often less integrated than end systems. One implication of this is reduced performance for certain functions (e.g., communication).

**State of the Art**

The increased scale and reduced integration of the cluster environment make the implementation of certain services more difficult and also introduce a need for new services not required in a single end system. The result tends to be either significantly reduced performance (and hence range of applications) or software architectures that modify and/or extend end-system operating systems in significant ways.

We use the problem of high-performance parallel execution to illustrate the types of issues that can arise when we seek to provide familiar end-system services in a cluster environment. In a single (multiprocessor) end system, high-performance parallel execution is typically achieved either by using specialized communication libraries such as MPI or by creating multiple threads that communicate by reading and writing a shared address space.

Both message-passing and shared-memory programming models can be implemented in a cluster. Message passing is straightforward to implement, since the commodity systems from which clusters are constructed typically support at least TCP/IP as a communication protocol. Shared memory requires additional effort: in an end system, hardware mechanisms ensure a uniform address space for all threads, but in a cluster, we are dealing with multiple address spaces. One approach to this problem is to implement a logical shared memory by providing software mechanisms for translating between local and global addresses, ensuring coherency between different versions of data, and so forth. A variety of such distributed shared-memory systems exist, varying according to the level at which sharing is permitted [76], [24], [53].

In low-performance environments, the cluster developer's job is done at this point; message-passing and DSM systems can be run as user-level programs that use conventional communication protocols and mechanisms (e.g., TCP/IP) for interprocessor communication. However, if performance is important, considerable additional development effort may be required. Conventional network protocols are orders of magnitude slower than intra-end-system communication operations. Low-latency, high-bandwidth inter-end-system communication can require modifications to the protocols used for communication, the operating system's treatment of network interfaces, or even the network interface hardware [70], [56].

The cluster developer who is concerned with parallel performance must also address the problem of coscheduling. There is little point in communicating extremely rapidly to a remote process that must be scheduled before it can respond. Coscheduling refers to techniques that seek to schedule simultaneously the processes constituting a computation on different processors [23], [63]. In certain highly integrated parallel computers, coscheduling is achieved by using a batch scheduler: processors are space shared, so that only one computation uses a processor at a time. Alternatively, the schedulers on the different systems can communicate, or the application itself can guide the local scheduling process to increase the likelihood that processes will be coscheduled [3], [14].

To summarize the points illustrated by this example: in clusters, the implementation of services taken for granted in end systems can require new approaches to the implementation of existing services (e.g., interprocess communication) and the development of new services (e.g., DSM and coscheduling). The complexity of the new approaches and services, as well as the number of modifications required to the commodity technologies from which clusters are constructed, tends to increase proportionally with performance requirements.

We can paint a similar picture in other areas, such as process creation, process control, and I/O.

Experience shows that familiar services can be extended to the cluster environment without too much difficulty, especially if performance is not critical; the more sophisticated cluster systems provide transparent mechanisms for allocating resources, creating processes, controlling processes, accessing files, and so forth, that work regardless of a program's location within the cluster. However, when performance is critical, new implementation techniques, low-level services, and high-level interfaces can be required [65], [25].

**Future Directions**

Cluster architectures are evolving in response to three pressures:

1. Performance requirements motivate increased integration and hence operating system and hardware modifications (for example, to support fast communications).

2. Changed operational parameters introduce a need for new operating system and user-level services, such as coscheduling.

3. Economic pressures encourage a continued focus on commodity technologies, at the expense of decreased integration and hence performance and services.

It seems likely that, in the medium term, software architectures for clusters will converge with those for end systems, as end-system architectures address issues of network operation and scale.

## 5.4  Intranets

The third class of systems that we consider is the intranet, a grid comprising a potentially large number of resources that nevertheless belong to a single organization. Like a cluster, an intranet can assume centralized administrative control and hence a high degree of coordination among resources. The three principal complicating factors that an intranet introduces are as follows:

1. *Heterogeneity:* The end systems and networks used in an intranet are almost certainly of different types and capabilities. We cannot assume a single system image across all end systems.

2. *Separate administration:* Individual systems will be separately administered; this feature introduces additional heterogeneity and the need to negotiate potentially conflicting policies.

3. *Lack of global knowledge:* A consequence of the first two factors, and the increased number of end systems, is that it is not possible, in general, for any one person or computation to have accurate global knowledge of system structure or state.

**State of the Art**

The software technologies employed in intranets focus primarily on the problems of physical and administrative heterogeneity. The result is typically a simpler, less tightly integrated set of services than in a typical cluster. Commonly, the services that are provided are concerned primarily with the sharing of data (e.g., distributed file systems, databases, Web services) or with providing access to specialized services, rather than with supporting the coordinated use of multiple resources. Access to nonlocal resources often requires the use of simple, high-level interfaces designed for "arm's-length" operation in environments in which every operation may involve authentication, format conversions, error checking, and accounting. Nevertheless, centralized administrative control does mean that a certain degree of uniformity of mechanism and interface can be achieved; for example, all machines may be required to run a specific distributed file system or batch scheduler, or may be placed behind a firewall, hence simplifying security solutions.

Software architectures commonly used in intranets include the Distributed Computing Environment (DCE), DCOM, and CORBA. In these systems, programs typically do not allocate resources and create processes explicitly, but rather connect to established "services" that encapsulate hardware resources or provide defined computational services. Interactions occur via remote procedure call [33] or remote method invocation [55], [36], models designed for situations in which the parties involved have little knowledge of each other. Communications occur via standardized protocols (typically layered on TCP/IP) that are designed for portability rather than high performance. In larger intranets, particularly those used for mission-critical applications, reliable group communication protocols such as those implemented by ISIS [7] and Totem [46] can be used to deal with failure by ordering the occurrence of events within the system.

The limited centralized control provided by a parent organization can allow the deployment of distributed queuing systems such as Load Sharing Facility (LSF), Codine, or Condor, hence providing uniform access to compute resources. Such systems provide some support for remote management of computation, for example, by distributing a limited range of signals to processes through local servers and a logical signal distribution network. However, issues of security, payment mechanisms, and policy often prevent these solutions from scaling to large intranets.

In a similar fashion, uniform access to data resources can be provided by means of wide area file system technology (such as DFS), distributed database technology, or remote database access (such as SQL servers). High-performance, parallel access to data resources can be provided by more specialized systems such as the High Performance Storage System [73]. In these cases, the interfaces presented to the application would be the same as those provided in the cluster environment.

The greater heterogeneity, scale, and distribution of the intranet environment also introduce the need for services that are not needed in clusters. For example, resource discovery mechanisms may be needed to support the discovery of the name, location, and other characteristics of resources currently available on the network. A reduced level of trust and greater exposure to external threats may motivate the use of more sophisticated security technologies. Here, we can once again exploit the limited centralized control that a parent organization can offer. Solutions such as Kerberos [51] can be mandated and integrated into the computational model, providing a unified authentication structure throughout the intranet.

**Future Directions**

Existing intranet technologies do a reasonable job of projecting a subset of familiar programming models and services (procedure calls, file systems, etc.) into an environment of greater complexity and physical scale, but are inadequate for performance-driven applications. We expect future developments to overcome these difficulties by extending lighter-weight interaction models originally developed within clusters into the more complex intranet environment, and by developing specialized performance-oriented interfaces to various services.

### 5.5 Internets

The final class of systems that we consider is also the most challenging on which to perform network computing—internetworked systems that span multiple organizations. Like intranets, internets tend to be large and heterogeneous. The three principal additional complicating factors that an internet introduces are as follows:

1. *Lack of centralized control:* There is no central authority to enforce operational policies or to ensure resource quality, and so we see wide variation in both policy and quality.

2. *Geographical distribution:* Internets typically link resources that are geographically widely distributed. This distribution leads to network performance characteristics significantly different from those in local area or metropolitan area networks of clusters and intranets. Not only does latency scale linearly with distance, but bisection bandwidth arguments [18], [26] suggest that accessible bandwidth tends to decline linearly with distance, as a result of increased competition for long-haul links.

3. *International issues:* If a grid extends across international borders, export controls may constrain the technologies that can be used for security, and so on.

**State of the Art**

The internet environment's scale and lack of central control have so far prevented the successful widespread deployment of grid services. Approaches that are effective in intranets often break down because of the increased scale and lack of centralized management. The set of assumptions that one user or resource can make about another is reduced yet further, a situation that can lead to a need for implementation techniques based on discovery and negotiation.

We use two examples to show how the internet environment can require new approaches. We first consider security. In an intranet, it can be reasonable to assume that every user has a preestablished trust relationship with every resource that he wishes to access. In the more open internet environment, this assumption becomes intractable because of the sheer number of potential process-to-resource relationships. This problem is accentuated by the dynamic and transient nature of computation, which makes any explicit representation of these relationships infeasible. Free-flowing interaction between computations and resources requires more dynamic approaches to authentication and access control. One potential solution is to introduce the notion of delegation of trust into security relationships; that is, we introduce mechanisms that allow an organization A to trust a user U because user U is trusted by a second organization B, with which A has a formal relationship. However, the development of such mechanisms remains a research problem.

As a second example, we consider the problem of coscheduling. In an intranet, it can be reasonable to assume that all resources run a single scheduler, whether a commercial system such as LSF or a research system such as Condor. Hence, it may be feasible to provide coscheduling facilities in support of applications that need to run on multiple resources at once. In an internet, we cannot rely on the existence of a common scheduling infrastructure. In this environment, coscheduling requires that a grid application (or scheduling service acting for an application) obtain knowledge of the scheduling policies that apply on different resources and influence the schedule either directly through an external scheduling API or indirectly via some other means [16].

**Future Directions**

Future development of grid technologies for internet environments will involve the development of more sophisticated grid services and the gradual evolution of the services provided at end systems in support of those services. There is little consensus on the shape of the grid architectures that will emerge as a result of this process, but both commercial technologies and research projects point to interesting potential directions. Three of these directions—commodity technologies, Legion, and Globus—are explored in detail in later chapters. We note their key characteristics here but avoid discussion of their relative merits. There is as yet too little experience in their use for such discussion to be meaningful.

The commodity approach to grid architecture adopts as the basis for grid development the vast range of commodity technologies that are emerging at present, driven by the success of the Internet and Web and by the demands of electronic information delivery and commerce. These technologies are being used to construct three-tier architectures, in which middle-tier application servers mediate between sophisticated back-end services and potentially simple front ends. Grid applications are supported in this environment by means of specialized high-performance back-end and application servers.

The Legion approach to grid architecture seeks to use object-oriented design techniques to simplify the definition, deployment, application, and long-term evolution of grid components. Hence, the Legion architecture defines a complete object model that includes abstractions of compute resources called *host objects,* abstractions of storage systems called *data vault objects,* and a variety of other object classes. Users can use inheritance and other object-oriented techniques to specialize the behavior of these objects to their own particular needs, as well as develop new objects.

The Globus approach to grid architecture is based on two assumptions:

1. Grid architectures should provide basic services, but not prescribe particular programming models or higher-level architectures.

2. Grid applications require services beyond those provided by today's commodity technologies.

Hence, the focus is on defining a "toolkit" of low-level services for security, communication, resource location, resource allocation, process management, and data access. These services are then used to implement higher-level services, tools, and programming models.

In addition, hybrids of these different architectural approaches are possible and will almost certainly be addressed; for example, a commodity three-tier system might use Globus services for its back end.

A wide range of other projects are exploring technologies of potential relevance to computational grids, for example, WebOS [67], Charlotte [6], UFO [2], ATLAS [5], Javelin [15], Popcorn [10], and Globe [69].

## 6 Research Challenges

What problems must be solved to enable grid development? In preceding sections, we outlined what we expect grids to look like and how we expect them to be used. In doing so, we tried to be as concrete as possible, with the goal of providing at least a plausible view of the future. However, there are certainly many challenges to be overcome before grids can be used as easily and flexibly as we have described. In this section, we summarize the nature of these challenges, most of which are discussed in much greater detail in the chapters that follow.

### 6.1 The Nature of Applications

Early metacomputing experiments provide useful clues regarding the nature of the applications that will motivate and drive early grid development. However, history also tells us that dramatic changes in capabilities such as those discussed here are likely to lead to radically new ways of using computers— ways as yet unimagined. Research is required to explore the bounds of what is possible, both within those scientific and engineering domains in which metacomputing has traditionally been applied, and in other areas such as business, art, and entertainment.

## 6.2 Programming Models and Tools

As noted in Section 4, grid environments will require a rethinking of existing programming models and, most likely, new thinking about novel models more suitable for the specific characteristics of grid applications and environments. Within individual applications, new techniques are required for expressing advanced algorithms, for mapping those algorithms onto complex grid architectures, for translating user performance requirements into system resource requirements, and for adapting to changes in underlying system structure and state. Increased application and system complexity increases the importance of code reuse, and so techniques for the construction and composition of grid-enabled software components will be important. Another significant challenge is to provide tools that allow programmers to understand and explain program behavior and performance.

## 6.3 System Architecture

The software systems that support grid applications must satisfy a variety of potentially conflicting requirements. A need for broad deployment implies that these systems must be simple and place minimal demands on local sites. At the same time, the need to achieve a wide variety of complex, performance-sensitive applications implies that these systems must provide a range of potentially sophisticated services. Other complicating factors include the need for scalability and evolution to future systems and services. It seems likely that new approaches to software architecture will be needed to meet these requirements—approaches that do not appear to be satisfied by existing Internet, distributed computing, or parallel computing technologies.

## 6.4 Algorithms and Problem-Solving Methods

Grid environments differ substantially from conventional uniprocessor and parallel computing systems in their performance, cost, reliability, and security characteristics. These new characteristics will undoubtedly motivate the development of new classes of problem-solving methods and algorithms. Latency-tolerant and fault-tolerant solution strategies represent one important area in which research is required [5], [6], [10]. Highly concurrent and speculative execution techniques may be appropriate in environments where many more resources are available than at present.

## 6.5 Resource Management

A defining feature of computational grids is that they involve sharing of networks, computers, and other resources. This sharing introduces challenging resource management problems that are beyond the state of the art in a variety of areas. Many of the applications described in later chapters need to meet stringent end-to-end performance requirements across multiple computational resources connected by heterogeneous, shared networks. To meet these requirements, we must provide improved methods for specifying application-level requirements, for translating these requirements into computational resources and network-level quality-of-service parameters, and for arbitrating between conflicting demands.

## 6.6 Security

Sharing also introduces challenging security problems. Traditional network security research has focused primarily on two-party client-server interactions with relatively low performance requirements. Grid applications frequently involve many more entities, impose stringent performance requirements, and involve more complex activities such as collective operations and the downloading of code. In larger grids, issues that arise in electronic markets become important. Users may require assurance and licensing mechanisms that can provide guarantees (backed by financial obligations) that services behave as advertised [37].

### 6.7 Instrumentation and Performance Analysis

The complexity of grid environments and the performance complexity of many grid applications make techniques for collecting, analyzing, and explaining performance data of critical importance. Depending on the application and computing environment, poor performance as perceived by a user can be due to any one or a combination of many factors: an inappropriate algorithm, poor load balancing, inappropriate choice of communication protocol, contention for resources, or a faulty router. Significant advances in instrumentation, measurement, and analysis are required if we are to be able to relate subtle performance problems in the complex environments of future grids to appropriate application and system characteristics.

### 6.8 End Systems

Grids also have implications for the end systems from which they are constructed. Today's end systems are relatively small and are connected to networks by interfaces and with operating system mechanisms originally developed for reading and writing slow disks. Grids require that this model evolve in two dimensions. First, by increasing demand for high-performance networking, grid systems will motivate new approaches to operating system and network interface design in which networks are integrated with computers and operating systems at a more fundamental level than is the case today. Second, by developing new applications for networked computers, grids will accelerate local integration and hence increase the size and complexity of the end systems from which they are constructed.

### 6.9 Network Protocols and Infrastructure

Grid applications can be expected to have significant implications for future network protocols and hardware technologies. Mainstream developments in networking, particularly in the Internet community, have focused on best-effort service for large numbers of relatively low-bandwidth flows. Many of the future grid applications discussed in this book require both high bandwidths and stringent performance assurances. Meeting these requirements will require major advances in the technologies used to transport, switch, route, and manage network flows.

## 7 Summary

This chapter has provided a high-level view of the expected purpose, shape, and architecture of future grid systems and, in the process, sketched a road map for more detailed technical discussion in subsequent chapters. The discussion was structured in terms of six questions.

*Why do we need computational grids?* We explained how grids can enhance human creativity by, for example, increasing the aggregate and peak computational performance available to important applications and allowing the coupling of geographically separated people and computers to support collaborative engineering. We also discussed how such applications motivate our requirement for a software and hardware infrastructure able to provide dependable, consistent, and pervasive access to high-end computational capabilities.

*What types of applications will grids be used for?* We described five classes of grid applications: distributed supercomputing, in which many grid resources are used to solve very large problems; high throughput, in which grid resources are used to solve large numbers of small tasks; on demand, in which grids are used to meet peak needs for computational resources; data intensive, in which the focus is on coupling distributed data resources; and collaborative, in which grids are used to connect people.

*Who will use grids?* We examined the shape and concerns of four grid communities, each supporting a different type of grid: a national grid, serving a national government; a private grid, serving a health maintenance organization; a virtual grid, serving a scientific collaboratory; and a public grid, supporting a market for computational services.

*How will grids be used?* We analyzed the requirements of five classes of users for grid tools and services, distinguishing between the needs and concerns of end users, application developers, tool developers, grid developers, and system managers.

*What is involved in building a grid?* We discussed potential approaches to grid architecture, distinguishing between the differing concerns that arise and technologies that have been developed within individual end systems, clusters, intranets, and internets.

*What problems must be solved to enable grid development?* We provided a brief review of the research challenges that remain to be addressed before grids can be constructed and used on a large scale.

## Further Reading

For more information on the topics covered in this chapter, see *www.mkp.com/grids* and also the following references:

- A series of books published by the Corporation for National Research Initiatives [29], [30], [31], [28] review and draw lessons from other large-scale infrastructures, such as the electric power grid, telecommunications network, and banking system.

- Catlett and Smarr's original paper on metacomputing [13] provides an early vision of how high-performance distributed computing can change the way in which scientists and engineers use computing.

- Papers in a 1996 special issue of the *International Journal of Supercomputer Applications* [19] describe the architecture and selected applications of the I-WAY metacomputing experiment.

- Papers in a 1997 special issue of the *Communications of the ACM* [62] describe plans for a National Technology Grid.

- Several reports by the National Research Council touch upon issues relevant to grids [49], [50], [48].

- Birman and van Renesse [8] discuss the challenges that we face in achieving reliability in grid applications.